\documentclass{article}
\usepackage{spconf,amsmath,amssymb,graphicx,hyperref}
\usepackage{multirow}
\usepackage{makecell}
\usepackage{booktabs}
\usepackage{CJKutf8}
\title{Group Relative Policy Optimization for Text-to-Speech with Large Language Models}
\name{Chang Liu$^1$, Ya-Jun Hu$^2$, Ying-Ying Gao$^3$, Shi-Lei Zhang$^3$,  Zhen-Hua Ling$^{1*}$\thanks{$^*$Corresponding author.}}
\address{$^1$National Engineering Research Center of Speech and Language Information Processing, \\ University of Science and Technology of China, Hefei, China \\
$^2$iFLYTEK Research, Hefei, China \\
$^3$Jiutian Artificial Intelligence Research Institute, China Mobile} 
\begin{document}
\ninept
\maketitle
\begin{abstract}
This paper proposes a GRPO-based approach to enhance the performance of large language model (LLM)-based text-to-speech (TTS) models by deriving rewards from an off-the-shelf automatic speech recognition (ASR) model. Compared to previous reinforcement learning methods for LLM-based TTS, our method requires no dedicated model for reward computation or training. Moreover, we design a composite reward function that combines character error rate (CER) with negative log-likelihood (NLL) obtained from the ASR model, providing more informative and accurate reward signals. We apply GRPO fine-tuning to pre-trained LLM-based TTS models and evaluate their zero-shot TTS performance. Experimental results show that the proposed method substantially improves both the intelligibility and naturalness of synthesized speech. Ablation studies and further analyses confirm the effectiveness of integrating the two reward components.
\end{abstract}
\vspace{-0.2cm}
\section{Introduction}
\label{sec:intro}
\vspace{-0.2cm}

In recent years, large language models (LLM)-based TTS models have become the mainstream approach for speech synthesis due to their ability to generate highly natural-sounding speech and their powerful zero-shot cloning capabilities \cite{valle,lauragpt,cosyvoice,cosyvoice2,cosyvoice3,llasa,sparktts}. 
Given only a few seconds of prompt speech, these models can effectively capture speaker timbre, prosody, and speaking style, and synthesize speech corresponding to arbitrary text inputs. 

LLM-based TTS models can be broadly divided into two categories. 
The first uses large language models to model acoustic speech token extracted by speech codec models \cite{valle,lauragpt}. In this paradigm, the predicted speech token can be simply converted to speech waveform through the decoder of codec models. However, acoustic tokens contain rich information and lack alignment with text, which can easily lead to synthesized speech content inconsistency with the text, or even synthesis failure.

The second category models semantic speech tokens within the LLM, followed by a non-autoregressive flow-matching model to supplement acoustic details beyond the semantic information \cite{seedtts,cosyvoice}. 
The semantic tokens are usually extracted from an automatic speech recognition (ASR) models with a vector quantization layers, providing better alignment with text. 
Therefore, these models exhibit fewer synthesis failures, while the flow-matching module further improves the acoustic fidelity and naturalness of the generated speech. They usually outperform the first type.

While autoregressive sampling in LLM-based TTS models enables the generation of diverse and prosody-consistent speech, it also causes the model to sometimes generate speech that does not align with human preferences.
Therefore, some studies apply reinforcement learning (RL) to fine-tune LLM-based TTS models, thereby achieving better performance.
Seed-TTS \cite{seedtts} proposed to integrate RL using word error rate (WER) and speaker similarity (SIM) as rewards within the Proximal Policy Optimization (PPO) \cite{ppo} and REINFORCE \cite{REINFORCE} frameworks.
However, this approach requires maintaining and training multiple models simultaneously, which results in a complex and unstable training process.
To address this, many studies adopt Direct Preference Optimization (DPO) \cite{dpo} and its variants to enhance LLM-based TTS models \cite{emodpo,dpollmtts,koeltts}.
Although DPO does not require additional models, it heavily relies on high-quality paired preference data, making it sensitive to noisy or inconsistent annotations and costly to scale. Moreover, it offers limited fine-grained control over the reward function and shows restricted generalization ability \cite{limitdpo,isdposuper,DPOvsGRPO}.
DiffRO \cite{cosyvoice3,diffro} proposes a differentiable reward framework for fine-tuning speech token LLMs in a supervised manner. Specifically, Gumbel-Softmax is employed to approximate the sampling of output tokens at each step, while an ASR-style token-to-text model is built to evaluate the negative log-likelihood (NLL) between the generated speech tokens and the corresponding ground-truth transcripts, which serves as the reward signal. Owing to the differentiability of Gumbel-Softmax, the optimization can be directly formulated as minimizing the NLL. Nevertheless, this approach requires pre-training a token-to-text model, which incurs additional computational and data costs.

However, no prior study has applied Group Relative Policy Optimization (GRPO) \cite{deepseekmath} to fine-tune LLM-based TTS models. GRPO eliminates the value model used in PPO, thereby reducing resource consumption and training complexity. Unlike DPO, it does not require large amounts of paired positive and negative preference data in advance and can instead be trained solely with text data. In addition, by applying group-wise normalization, GRPO keeps the advantage function within a stable range, which contributes to stabilizing the training process.

In this paper, we propose a GRPO-based fine-tuning approach applicable to both categories of LLM-based TTS models introduced earlier. Our method leverages an off-the-shelf ASR model to directly derive rewards from generated speech waveforms and compute group-relative advantages, thereby substantially simplifying and stabilizing the training pipeline. Consequently, it obviates the need for a dedicated token-to-text model as required by DiffRO \cite{diffro}.
Additionally, we use a reward function that combines character error rate (CER) with negative log-likelihood (NLL), effectively balancing alignment accuracy and probabilistic confidence to achieve improved performance.

We fine-tune both categories of LLM-based TTS models using the proposed method and evaluate their zero-shot performance. Experimental results show that our approach significantly enhances both the semantic consistency and naturalness of the synthesized speech. Moreover, both ablation studies and additional analyses demonstrate the effectiveness of integrating the two reward components. Audio demos, codes and models are available at \url{https://ryuclc.github.io/LLM-TTS-GRPO}.

\vspace{-0.2cm}
\section{Proposed Method}
\label{sec:method}

\vspace{-0.2cm}
\subsection{Reward function}
\label{ssec:reward}
\vspace{-0.2cm}

This paper focuses on enhancing semantic consistency and proposes a reward function applicable to both categories of LLM-based TTS models. Specifically, given the reconstructed speech waveform from the LLM outputs, rewards are computed using an off-the-shelf ASR model combining character error rate (CER) and negative log-likelihood (NLL).

CER is a widely adopted metric for evaluating speech intelligibility. As an intuitive measure based on edit distance, it directly reflects the transcription accuracy of synthesized speech relative to the ground-truth text. Its simplicity and interpretability make CER a reliable indicator of how well the generated speech conveys the intended content.
However, relying solely on CER as the reward function may introduce several limitations. First, CER only measures surface-level transcription accuracy, and therefore ignores the ASR model’s confidence in its predictions. This may lead to ambiguous optimization signals, especially when different speech outputs yield the same CER but differ significantly in quality. Second, CER is insensitive to acoustic nuances such as prosody or fluency, which are essential for natural-sounding speech synthesis. Third, CER is discrete, which may result in sparse or unstable reward signals during reinforcement learning. 
% As a consequence, optimizing with CER alone risks overlooking subtle but important aspects of speech quality, potentially causing the model to converge to suboptimal solutions.

To address these issues, NLL is introduced as a complementary reward. NLL captures the probability distribution of the ASR model over the ground-truth tokens, providing a continuous, fine-grained signal that reflects the model’s confidence and sensitivity to subtle variations. By combining CER and NLL, the reward function balances objective intelligibility with model confidence, resulting in more stable training and higher-quality speech synthesis.

\begin{figure}[t]
	\centering
	\includegraphics[width=0.60\linewidth]{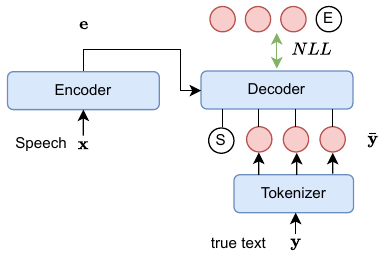}
        \vspace{-0.4cm}
	\caption{The process of obtaining NLL metric. \scalebox{1.1}{\textcircled{\raisebox{-0.8pt}{\footnotesize S}}} and \scalebox{1.1}{\textcircled{\raisebox{-0.8pt}{\footnotesize E}}} denote the ``start of transcribe" and ``end of transcribe" tokens respectively. The blue modules represent the ASR model.}
    \label{fig:nll}
    \vspace{-0.6cm}
\end{figure}

Formally, the ASR model transcribes input speech $\mathbf{x}$ into predicted text $\mathbf{\hat{y}}$, from which CER is computed against the ground-truth $\mathbf{y}$. For NLL, as shown in Figure~\ref{fig:nll}, the true text is tokenized as $\mathbf{\bar{y}} = (\bar{y}_1, \dots, \bar{y}_N)$, and passed through the ASR decoder to obtain logits. NLL is then calculated as:
\begin{equation}
\setlength\abovedisplayskip{3pt}
\setlength\belowdisplayskip{3pt}
NLL = - \sum_{n=1}^{N} \log P(\bar{y}_n | \bar{y}_{<n}, \mathbf{e}),
\end{equation}
where $\mathbf{e}$ is the encoder output, $\bar{y}_n$ is the $n$-th ground-truth token, and $P(\bar{y}_n | \bar{y}_{<n}, \mathbf{e})$ denotes the probability assigned by the ASR model. Lower NLL values indicate higher confidence and better alignment with the reference text.

\begin{figure*}[t]
	\centering
	\includegraphics[width=0.8\linewidth]{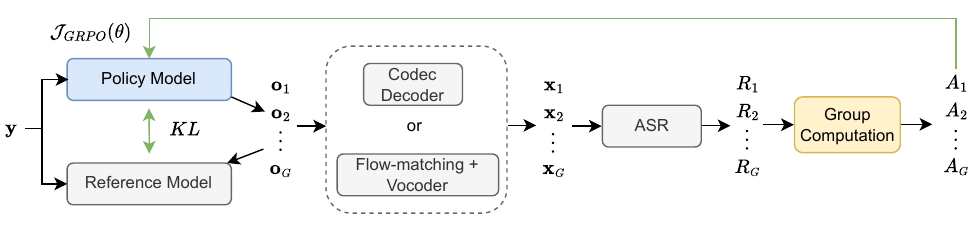}
        \vspace{-0.5cm}
	\caption{The process of GRPO fine-tuning. The gray modules are frozen.}
    \label{fig:grpo}
    \vspace{-0.6cm}
\end{figure*}

After obtaining the two metrics, we compute the CER reward and the NLL reward using the following formulas:
\begin{equation}
\setlength\abovedisplayskip{3pt}
\setlength\belowdisplayskip{3pt}
    R_{CER} = 1 - \tanh(\alpha_c \cdot CER),
\end{equation}
\begin{equation}
\setlength\abovedisplayskip{3pt}
\setlength\belowdisplayskip{3pt}
    R_{NLL} = \exp \left(-\frac{NLL}{\alpha_n}\right),
\end{equation}
where $\alpha_c$ and $\alpha_n$ control the sensitivity of the reward function to CER and NLL, respectively. Both formulas inherently map the metrics to the range $[0, 1]$.
The final reward function is obtained by combining the two components via a weighted harmonic mean:
\begin{equation}
\setlength\abovedisplayskip{3pt}
\setlength\belowdisplayskip{3pt}
R = \frac{\lambda_c + \lambda_n}{\dfrac{\lambda_c}{R_{CER}} + \dfrac{\lambda_n}{R_{NLL}}}.
\end{equation}
Here, $\lambda_c$ and $\lambda_n$ denote the weights assigned to CER and NLL rewards, respectively. Compared with the arithmetic mean, the harmonic mean is more sensitive to lower values, effectively penalizing extreme cases. For instance, if one metric performs poorly (e.g., a high CER), the overall reward is substantially reduced, preventing strong performance on one metric from masking deficiencies in the other.

\vspace{-0.4cm}
\subsection{Model fine-tuning with GRPO}
\label{ssec:grpo}
\vspace{-0.2cm}

In this work, we enhance pre-trained LLM-based TTS models using the GRPO algorithm \cite{deepseekmath}. The fine-tuning process is illustrated in Figure~\ref{fig:grpo}. The pre-trained speech token LLM serves as the policy model $\pi_\theta$, where $\theta$ denotes the trainable parameters. These parameters are also used to initialize the reference model $\pi_{ref}$, which remains frozen during fine-tuning.

Given an input text $\mathbf{y}$, the policy model autoregressively samples $G$ times, producing a set of outputs denoted as $O=\{\mathbf{o}_1, \mathbf{o}_2, \cdots, \mathbf{o}_G \}$. Here, $\mathbf{o}_i=[o_{i,1}, o_{i,2}, \cdots, o_{i,t}, \cdots ]$. The outputs are then used to compute the Kullback–Leibler (KL) divergence between the policy and reference models:
\begin{equation}
\setlength\abovedisplayskip{3pt}
\setlength\belowdisplayskip{3pt}
\begin{split}
\mathbb{D}_{KL}\left[ \pi_{\theta}||\pi_{ref} \right] = &  \frac{\pi_{ref}(o_{i,t}|\mathbf{y}, o_{i,<t})}{\pi_{\theta}(o_{i,t}|\mathbf{y}, o_{i,<t})} \\
&  - \log \frac{\pi_{ref}(o_{i,t}|\mathbf{y}, o_{i,<t})}{\pi_{\theta}(o_{i,t}|\mathbf{y}, o_{i,<t})} - 1.
\end{split}
\end{equation}
KL divergence acts as a penalty to prevent the policy model from deviating excessively from the reference, thereby maintaining training stability.

To obtain rewards, the inference outputs are converted into speech waveforms via a speech codec decoder or flow-matching with a vocoder. The resulting speech set $X=\{\mathbf{x}_1, \mathbf{x}_2, \cdots, \mathbf{x}_G \}$ is then fed into the ASR model to compute rewards $\mathbf{R}=[R_1, R_2, \cdots, R_G]$, as described in Section~\ref{ssec:reward}. Group-relative advantages are subsequently estimated as \cite{deepseekmath}:
\begin{equation}
\setlength\abovedisplayskip{3pt}
\setlength\belowdisplayskip{3pt}
A_i = \frac{R_i-mean(\mathbf{R})}{std(\mathbf{R})}
\end{equation}
Finally, the objective function is defined as follow:
\begin{equation}
\setlength\abovedisplayskip{3pt}
\setlength\belowdisplayskip{3pt}
\begin{split}
\mathcal{J}_{GRPO}(\theta) = \mathbb{E}[\mathbf{y} \sim \mathcal{D}, \{\mathbf{o}_{i}\}_{i=1}^G \sim \pi_\theta(O|\mathbf{y})]\frac{1}{G}\sum_{i=1}^{G} \\
\frac{1}{|\mathbf{o}_i|}\sum_{t=1}^{|\mathbf{o}_i|} \{ \frac{\pi_{\theta}(o_{i,t}|\mathbf{y}, o_{i,<t})}{\pi_{\theta_{old}}(o_{i,t}|\mathbf{y}, o_{i,<t})}A_{i}  - \beta \mathbb{D}_{KL} \left[ \pi_{\theta} || \pi_{ref} \right]\}.
\end{split}
\end{equation}
Here, $\mathcal{D}$ represents the dataset, $|\mathbf{o}_i|$ denotes the length of $\mathbf{o}_i$, and $\beta$ is the coefficient of the KL penalty. In GRPO, the old policy $\pi_{\theta_{old}}$ is usually initialized to the current policy $\pi_\theta$ at the start of each update and remains fixed during the gradient computation. Finally, the policy model parameters are updated by maximizing this objective function.

\vspace{-0.2cm}
\section{Experiments}
\label{sec:exp}
\vspace{-0.2cm}

\subsection{Experimental setup}
\label{ssec:setup}
\vspace{-0.2cm}
\subsubsection{Datasets and baseline models}
\vspace{-0.2cm}

To evaluate the effectiveness of the proposed GRPO method across different categories of LLM-based TTS models, we fine-tune two open-source baseline models with training codes: CosyVoice2 \cite{cosyvoice2} and Llasa-1B \cite{llasa}, which correspond to the two types of models introduced earlier. CosyVoice2 supports four languages: Chinese, English, Japanese, and Korean, whereas Llasa-1B supports only Chinese and English.

For GRPO fine-tuning, we randomly sampled 4,000 sentences from the Emilia dataset, covering Chinese, English, Japanese, and Korean. Chinese and English accounted for approximately 90\% of the total, with the remainder consisting of Japanese and Korean. The full set of 4,000 sentences was used to fine-tune CosyVoice2, while Llasa-1B was fine-tuned solely on the Chinese and English subsets.

\vspace{-0.4cm}
\subsubsection{GRPO setup}
\vspace{-0.2cm}

Our proposed method is denoted as GRPO-CER-NLL. In addition to comparing the baseline models with this approach, we conduct ablation studies to evaluate the contribution of each reward component. Specifically, GRPO-CER refers to fine-tuning using only the CER reward, while GRPO-NLL uses only the NLL reward. In our implementation, we adopt Whisper-large-v3 \cite{whisper} as the ASR model to obtain reward values.

For the reward function described in Section~\ref{ssec:reward}, both $\alpha_c$ and $\alpha_n$ are set to 3, while $\lambda_c$ and $\lambda_n$ are set to 0.6 and 0.4, respectively. During GRPO training, the KL penalty coefficient $\beta$ is set to 0.1, and the number of policy model inferences per group $G$ is set to 8. The learning rate is fixed at $1\times10^{-5}$.

\begin{table*}[t]
	\centering
	\caption{Objective evaluation results of zero-shot TTS using different models on objective sets, including content consistency (CER/WER) and speaker similarity (SIM).}
	\renewcommand\arraystretch{0.8}
        % \tabcolsep=0.12cm
	% \vspace{-0.2cm}
     \footnotesize
	\label{tab:zs-obj}
	\begin{tabular}{lcccccccc}
        \toprule
        \multirow{2}{*}[-1.3ex]{Model} & \multicolumn{2}{c}{zh} &  \multicolumn{2}{c}{en} & \multicolumn{2}{c}{ja} & \multicolumn{2}{c}{ko} \\
        \cmidrule(lr){2-3} \cmidrule(lr){4-5} \cmidrule(lr){6-7} \cmidrule(lr){8-9}
        & CER $\downarrow$ & SIM $\uparrow$ & CER $\downarrow$ & SIM $\uparrow$ & WER $\downarrow$ & SIM $\uparrow$ & CER $\downarrow$ & SIM $\uparrow$ \\
        \midrule
        \textbf{Human} & 1.33 & 0.755 & 2.10 & 0.734 & 8.53 & 0.708 & 7.43 & 0.716  \\
        \midrule
        \textbf{Llasa-1B} & 7.73 & 0.636 & 4.95 & 0.578 & - & - & - & - \\
        \textbf{+ GRPO-CER} & 1.72 & 0.672  & 2.61 & 0.580 & - & - & - & - \\
        \textbf{+ GRPO-NLL} & 1.05 & 0.674  & 2.49 & 0.581 & - & - & - & - \\
        \textbf{+ GRPO-CER-NLL} & 1.30 & 0.669  & 2.17 & 0.580 & - & - & - & - \\
        \midrule
        \textbf{CosyVoice2} & 1.41 & 0.753 & 2.46 & 0.655 & 12.45 & 0.635 & 8.58 & 0.670 \\
        \textbf{+ GRPO-CER} & 1.34 & 0.751  & 2.43 & 0.655 & 10.05 & 0.645 & 6.37 & 0.677 \\
        \textbf{+ GRPO-NLL} & 0.98 & 0.753  & 2.36 & 0.659 & 9.36 & 0.662 & 6.59 & 0.682 \\
        \textbf{+ GRPO-CER-NLL} & 1.07 & 0.753  & 2.30 & 0.659 & 9.09 & 0.656 & 6.16 & 0.680 \\
        \bottomrule
    \end{tabular}
		\vspace{-0.6cm}
\end{table*}

\begin{table}[!t]
	\centering
    \vspace{-0.3cm}
	\caption{MOS results of zero-shot TTS using different models on subjective sets with 95\% confidence intervals.}
	\renewcommand\arraystretch{0.8}
        \tabcolsep=0.08cm
        \footnotesize
	% \vspace{-0.2cm}
	\label{tab:zs-mos}
	\begin{tabular}{lcccc}
        \toprule
        Model & zh & en & ja & ko \\
        \midrule
        \textbf{CosyVoice2} & 4.42 $\pm$ 0.05 & 4.22 $\pm$ 0.06 & 4.10 $\pm$ 0.08 & 4.18 $\pm$ 0.08  \\
        \textbf{+ GRPO-CER} & 4.44 $\pm$ 0.06 & 4.26 $\pm$ 0.07 & 4.15 $\pm$ 0.08 & 4.23 $\pm$ 0.08  \\
        \textbf{+ GRPO-NLL} & 4.52 $\pm$ 0.05 & 4.31 $\pm$ 0.06 & 4.21 $\pm$ 0.08 & 4.24 $\pm$ 0.08  \\
        \textbf{+ GRPO-CER-NLL} & 4.58 $\pm$ 0.05 & 4.43 $\pm$ 0.06 & 4.29 $\pm$ 0.08 & 4.30 $\pm$ 0.08  \\
        \bottomrule
    \end{tabular}
		\vspace{-0.4cm}
\end{table}

\vspace{-0.3cm}
\subsection{Experimental results}
\label{ssec:result}
\vspace{-0.2cm}

\subsubsection{Evaluation sets}
\vspace{-0.2cm}
To evaluate the performance of zero-shot speech synthesis across different TTS models, we prepared both objective and subjective test sets. For Chinese (zh) and English (en), we adopted the open-source benchmarks provided by seed-tts-eval \cite{seedtts}, containing 2,020 and 1,088 samples, respectively. For Japanese (ja) and Korean (ko), we additionally collected 1,000 test samples per language from the Common Voice dataset \cite{commonvoice}. Furthermore, the subjective set is used for listening tests, as it includes high expressiveness and diverse styles. We collect approximately 100 samples per language.

\vspace{-0.3cm}
\subsubsection{Objective evaluation}
\vspace{-0.2cm}

We employed two objective metrics for evaluation:
1) content consistency (CER/WER),
which is used to assess speech intelligibility.For Chinese, we used Paraformer-zh \cite{funasr}, while for the other languages we used Whisper-large-v3 \cite{whisper} to transcribe the generated speech. 
word error rate (WER) was calculated for English, and character error rate (CER) for the remaining languages. For each subset, the average CER/WER was computed by dividing the total edit distance by the total reference length.
2) speaker similarity (SIM), which evaluates how closely the synthesized speech matches the reference speaker. It is calculated as the cosine similarity between the speaker embeddings of the generated and reference speech. The embeddings were extracted using a speaker verification model fine-tuned on WavLM \cite{wavlm-sv}.

The objective evaluation results of zero-shot TTS are presented in Table~\ref{tab:zs-obj}. As previously discussed, Llasa-1B \cite{llasa} relies on acoustic speech tokens, which lack alignment with the text, whereas CosyVoice2 \cite{cosyvoice2} utilizes semantic tokens within an LLM, followed by a flow-matching model to refine acoustic details. Consequently, \textbf{CosyVoice2} achieves significantly better performance than \textbf{Llasa-1B} on both metrics for Chinese and English.
Furthermore, \textbf{GRPO-CER-NLL} consistently improves CER/WER over both baseline models, demonstrating that our proposed method effectively enhances the semantic consistency of speech generated by LLM-based TTS models. Regarding speaker similarity (SIM), \textbf{GRPO-CER-NLL} is comparable to the baseline models, except in cases where the baseline exhibits extremely high CER/WER. In such cases, excessive pronunciation errors likely lead the WavLM-based speaker verification model to produce unreliable similarity scores.
Compared to the ablated models, \textbf{GRPO-CER-NLL} shows overall superior performance across different subsets, except for slightly lower performance than \textbf{GRPO-NLL} on Chinese. This highlights the complementary role of NLL as an additional reward signal in improving synthesis quality.

\vspace{-0.4cm}
\subsubsection{Subjective evaluation}
\vspace{-0.2cm}

To evaluate the impact of GRPO on the naturalness of synthesized speech, we conducted subjective listening tests. Specifically, we performed MOS evaluations on CosyVoice2 and its variants fine-tuned using different methods. For each language, 30 samples were randomly selected from the subjective evaluation sets for zero-shot synthesis. We then conducted listening tests by recruiting 10 native speakers for each language group, who rated the synthesized speech in terms of overall naturalness on a 1–5 scale with 0.5-point intervals.

The MOS results are presented in Table~\ref{tab:zs-mos}. The $p$-value of paired $t$-test was employed to measure the significance of the difference between two models. Across all languages, the proposed \textbf{GRPO-CER-NLL} model outperformed the baseline models ($p<0.05$), indicating that combining GRPO with the proposed reward function effectively enhances the performance of LLM-based TTS models. Additionally, \textbf{GRPO-CER} achieved significant higher MOS scores than the baseline across all languages ($p<0.05$) except Chinese ($p>0.05$), while \textbf{GRPO-NLL} outperformed \textbf{GRPO-CER} on all languages ($p<0.05$) except Korean ($p>0.05$). These results further demonstrate the benefits of integrating CER and NLL as complementary reward signals.

For Llasa-1B and its fine-tuned models, we also conduct subjective listening tests. However, the experimental results show that there are no significant improvements on naturalness with reinforcement learning. Therefore, we do not list the results here. The underlying reasons will be investigated in future work.

\vspace{-0.4cm}
\subsubsection{Analysis experiments}
\vspace{-0.2cm}

\begin{figure}[t]
	\centering
	\includegraphics[width=0.5\linewidth]{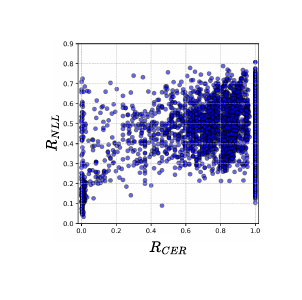}
        \vspace{-0.4cm}
	\caption{Scatter plot of $R_{CER}$ versus $R_{NLL}$.}
    \label{fig:corr}
    \vspace{-0.7cm}
\end{figure}

First, we examined the correlation between CER and NLL. We computed the rewards $R_{CER}$ and $R_{NLL}$ for synthesized speech generated by the two baseline models on the objective test sets across four languages. We observed that shorter speech segments tend to have lower $R_{NLL}$, as the ASR model exhibits lower confidence with limited context. To mitigate this, we removed excessively short sentences, resulting in approximately 8,000 data points. A scatter plot was then created with $R_{CER}$ on the horizontal axis and $R_{NLL}$ on the vertical axis as show in Figure~\ref{fig:corr}. The plot shows no strong correlation, and in most cases $R_{CER}$ equals 1 while $R_{NLL}$ provides additional discriminative information, demonstrating the effectiveness of incorporating NLL as a complementary reward. The Pearson correlation coefficient between $R_{CER}$ and $R_{NLL}$ was $r = 0.3371$, which, despite being statistically significant ($p < 0.05$), indicates a negligible linear relationship in practice.

\begin{CJK}{UTF8}{gbsn} % gbsn: 简体中文宋体, 可以换成 gkai (楷体) 等
\begin{figure}[t]
	\centering
	\includegraphics[width=0.9\linewidth]{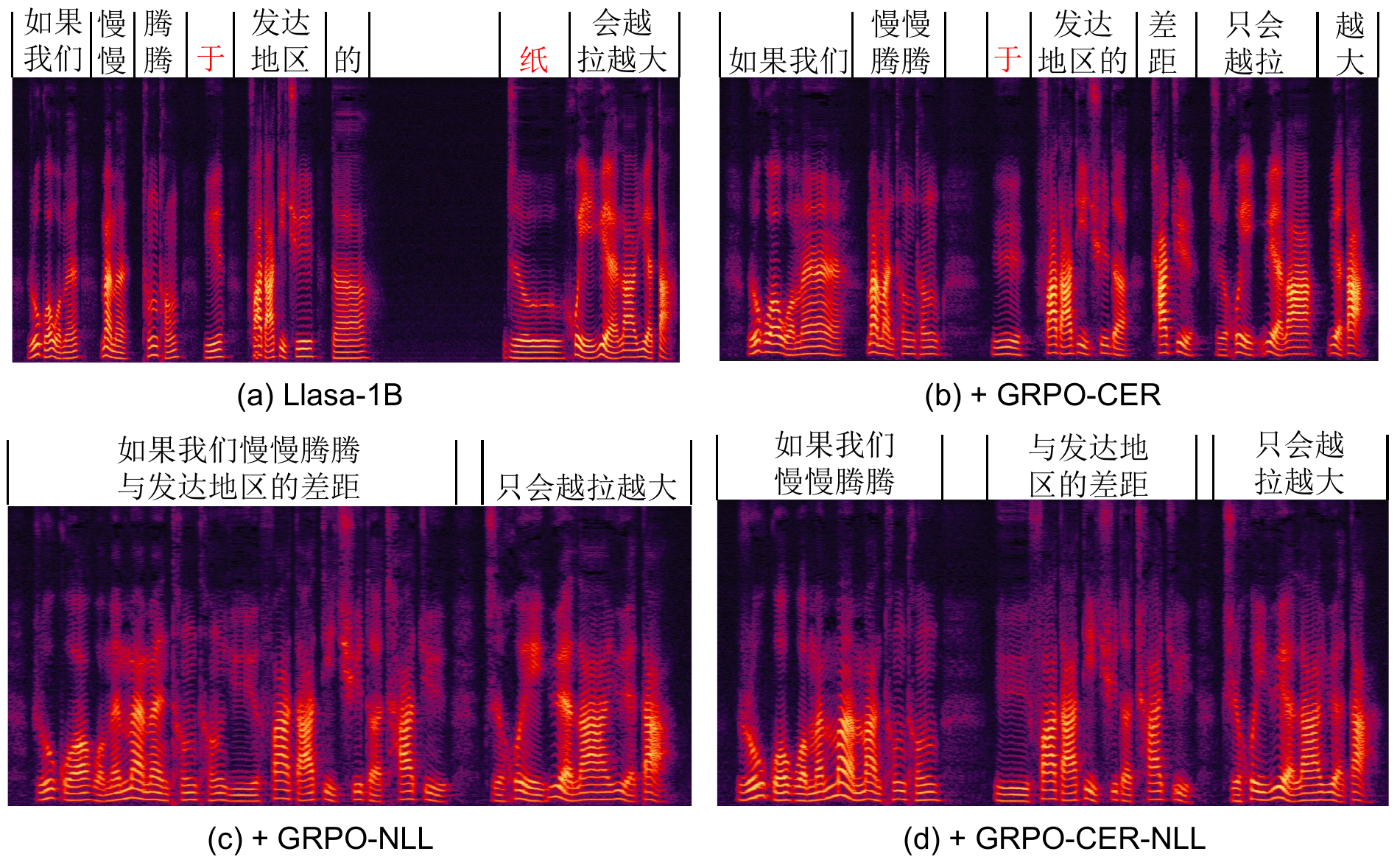}
        \vspace{-0.45cm}
	\caption{The spectrogram of synthesized speech by different models. The input text is ``如果我们慢慢腾腾，与发达地区的差距，只会越拉越大。"}
    \label{fig:analysis1}
    \vspace{-0.4cm}
\end{figure}
\end{CJK}

\begin{figure}[t]
	\centering
	\includegraphics[width=0.9\linewidth]{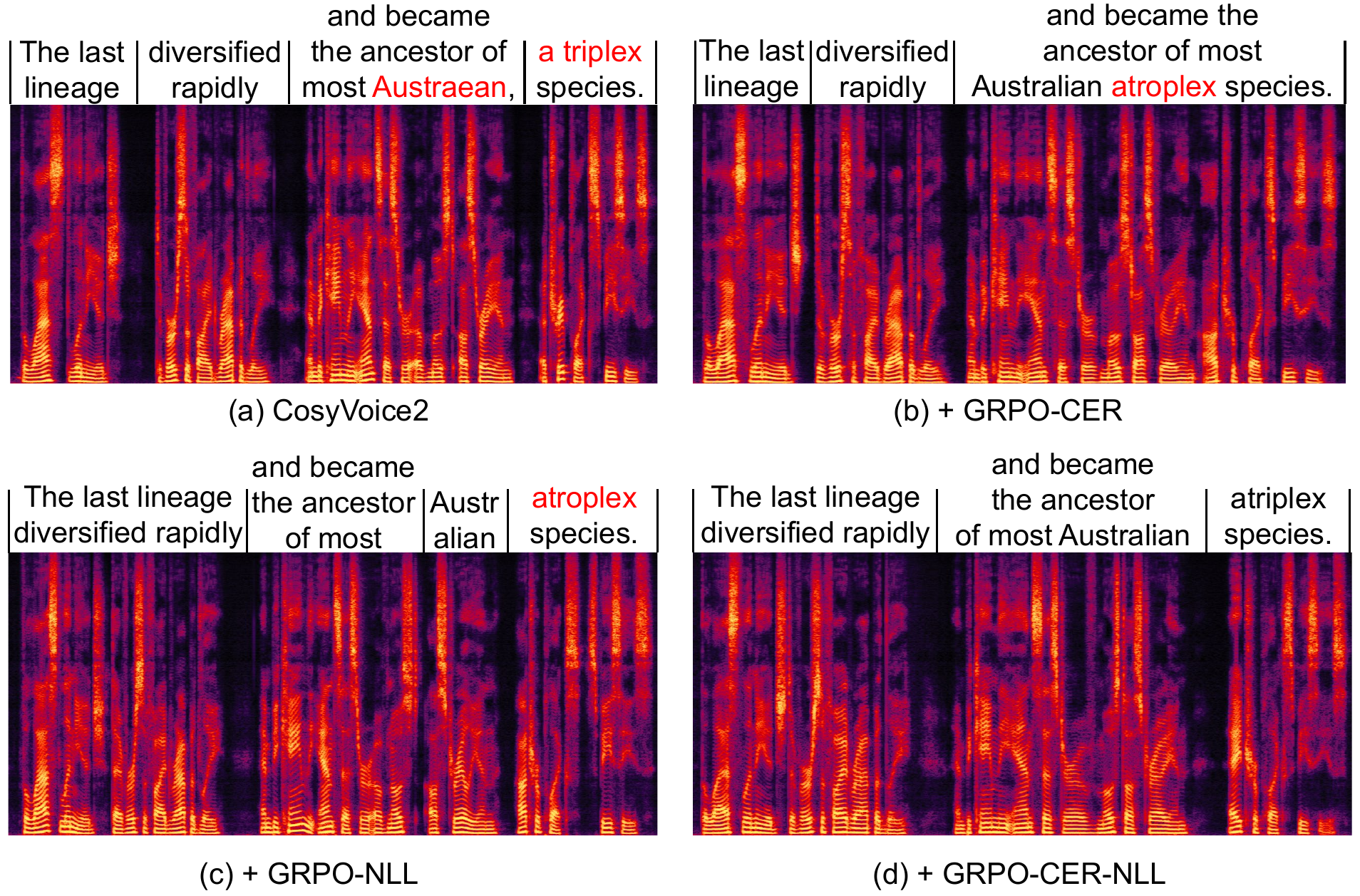}
        \vspace{-0.45cm}
	\caption{The spectrogram of synthesized speech by different models. The input text is ``The last lineage diversified rapidly, and became the ancestor of most Australian ``Atriplex" species.”}
    \label{fig:analysis2}
    \vspace{-0.6cm}
\end{figure}

\begin{CJK}{UTF8}{gbsn} % gbsn: 简体中文宋体, 可以换成 gkai (楷体) 等
Second, we compared pronunciations errors in speech generated by baseline models with those generated by models fine-tuned using our proposed GRPO-based methods.
Figure~\ref{fig:analysis1} shows a Chinese example synthesized by Llasa-1B and its variants through spectrogram. We can find that \textbf{Llasa-1B} produced pronunciation errors on ``于" and ``纸", and omitted ``差距" due to alignment issues. After GRPO fine-tuning, all variants corrected the omission of ``差距" and the error in ``纸". However, the \textbf{GRPO-CER} still mispronounced ``于", while \textbf{GRPO-NLL} and our proposed \textbf{GRPO-CER-NLL} produced fully correct pronunciations. In addition, \textbf{GRPO-CER-NLL} exhibited more natural prosodic breaks.

Besides, Figure~\ref{fig:analysis2} illustates the spectrogram of a English case generated by CosyVoice2 and its variants. They shows the similar results as in Figure~\ref{fig:analysis1}.
These analyses provide intuitive evidence for the superiority of our proposed method.
\end{CJK}

% \subsection{Analysis on reward function}
% \label{ssec:analysis}
\vspace{-0.4cm}
\section{Conclusion}
\label{sec:conclusion}
\vspace{-0.2cm}
In this paper, we proposed a GRPO approach to enhance the pre-trained LLM-based TTS models. This method adopts an off-the-shell ASR model to obtain rewards, reducing the reliance on extra models to make the training process simpler and more stable. Besides, this paper adopts a reward function which is derived from the CER and NLL. This reward combines objective intelligibility with ASR model confidence, resulting in improved semantic consistency and naturalness.

%\vfill
%\pagebreak

\bibliographystyle{IEEEbib}
\bibliography{refs}

\end{document}